\documentstyle[aps,epsf]{revtex}
\begin{document}


\title{Charge correlations and optical conductivity in weakly doped 
       antiferromagnets}
\author{Matthias Vojta and Klaus W. Becker}
\address{Institut f\"{u}r Theoretische Physik,
Technische Universit\"{a}t Dresden, D-01062 Dresden, \\ Germany}

\maketitle

\begin{abstract}
We investigate the dynamical charge-charge correlation function
and the optical conductivity in weakly 
doped antiferromagnets using Mori-Zwanzig projection technique.
The system is described by the two-dimensional $t$-$J$ model.
The arising matrix elements are evaluated within a cumulant
formalism which was recently applied to investigate magnetic properties
of weakly doped antiferromagnets. 
Within the present approach the ground state
consists of non-interacting hole quasiparticles.
Our spectra agree well with numerical results calculated via exact 
diagonalization techniques.
The method we employ enables us to 
explain the features present in the correlation functions.
We conclude that the charge dynamics at weak doping
is governed by transitions between excited states of spin-bag
quasiparticles.
\\
\end{abstract}



The dynamical properties of the normal state of the superconducting
cuprates are still not completely understood. A much discussed question
about strongly correlated electronic systems in two dimensions is
whether they can be described as Fermi liquids or not.
In Fermi liquids the low-lying spin and density excitations can be considered
as particle-hole excitations of fermionic quasiparticles.
In contrast, one-dimensional correlated systems show spin-charge
separation, i.e. spin and charge excitations are different elementary excitations,
and the "physical" electron can be interpreted as a composite object.

Theoretical progress in this field is mostly based on numerical
techniques, especially on exact diagonalization 
methods \cite{ToHoMae95,EdOhMae95,EdWroOh96}
and Lanczos calculations \cite{JaPre95}). However, the system
sizes presently accessible leave many problems unresolved. Meanwhile,
the spin response of doped antiferromagnets has been studied in a number
of analytical papers but only few authors have investigated the charge
density response function \cite{KhaHo96,ZeyKul96}.

In this paper we present analytical results based on projection technique 
calculations for dynamical charge-charge correlation functions
in weakly doped 2D antiferromagnets. The system is described 
by the 2D $t$-$J$ model:
\begin{equation}
H\, =\, - t \sum_{\langle ij\rangle \sigma}
      (\hat c^\dagger_{i\sigma} \hat c_{j\sigma} +
       \hat c^\dagger_{j\sigma} \hat c_{i\sigma})
    + J \sum_{\langle ij\rangle} \ ({\bf S}_i {\bf S}_j - {{n_i n_j} \over 4} )
\label{H_TJ}
\,.
\label{TJ_MOD}
\end{equation}
The symbol $\langle ij \rangle$ refers to a summation over pairs of 
nearest neighbors. The two antiferromagnetic
sublattices will be denoted by $\uparrow$ and $\downarrow$.
The electronic creation operators $\hat c_{i \sigma}^{\dagger}$
exclude double occupancies.

We study the Laplace transform of the time- and wave-vector dependent 
charge-charge correlation function at zero temperature defined by
\begin{equation}
G_\rho({\bf k},z) \,=\, 
  \langle\psi_0|\,\rho_{\bf k}^\dagger {1 \over {z-L}} \rho_{\bf k}^{\phantom\dagger}\,
  |\psi_0\rangle
\label{CCORR_DEF}
\end{equation}
Here,
$\rho_{\bf k} = \sum_{{\bf q},\sigma} 
c_{{\bf k+q},\sigma}^\dagger c_{{\bf q}\sigma}^{\phantom\dagger} =
\sum_{i\sigma} {\rm e}^{{\rm i}{\bf k}{\bf R}_i}
c_{i\sigma}^\dagger c_{i\sigma}^{\phantom\dagger}$
is the Fourier-transformed charge density operator.
$|\psi_0\rangle$ denotes the exact ground state of the system,
$z = \omega + {\rm i} \eta$ is the complex frequency variable.
The Liouville operator $L$ is a superoperator defined by $L A = [H,A]_{-}$
for any operator $A$.
At zero temperature the optical conductivity $\sigma(\omega)$ 
for frequencies $\omega > 0$ is related to the
charge response function $G_\rho({\bf k}, \omega)$ as follows:
\begin{equation}
{\rm Re}\,\sigma(\omega)\,=\, { {\rm e}^2 \over cN} \lim_{{\bf k} \rightarrow 0} 
\frac {\omega\,{\rm Im}\, G_\rho({\bf k}, \omega)} {{\bf k}^2}
\,
\label{OPTCON}
\end{equation}
where $N$ is the total particle number.
Eq. (\ref{OPTCON}) follows from the continuity relation
$L\rho_{\bf k} = {\bf k}\cdot{\bf j}_{\bf k}$ where
${\bf j}_{\bf k}$ is the charge current operator. 

In the following, we apply a projection technique \cite{Mori}
approach to determine the charge-density response function. The arising matrix 
elements are evaluated using a cumulant formalism \cite{BeckFul88,BeckBre90} 
which has been recently introduced to investigate ground-state 
properties of correlated electronic systems.
We are interested in calculating dynamical correlation functions for a
set of operators $B_{\nu}$
\begin{equation}
G_{\nu\mu}(z) \,=\, \langle\psi_0|\,\delta B_{\nu}^\dagger {1 \over {z-L}}
                   \delta B_{\mu}^{\phantom\dagger}\, |\psi_0\rangle
\label{ARBKORR}
\end{equation}
with $\delta B_{\nu} = B_{\nu} - \langle\psi_0| B_{\nu} |\psi_0\rangle$.
Using cumulants these
correlation functions can be rewritten as \cite{BeckBre90}
\begin{equation}
G_{\nu\mu}(z) \,=\,
  \langle\phi_0|\,\Omega^\dagger\,B_{\nu}^\dagger \left( {1 \over {z-L}}
  B_{\mu}^{\phantom\dagger} \right)^{\cdot} \, \Omega\,
  |\phi_0\rangle^c
\label{KUMKORR}
\end{equation}
The operator $\Omega$ has similarity to the so-called wave 
operator (or Moeller operator known from scattering theory). 
Within cumulants it transforms the ground state $|\phi_0\rangle$ of the
unperturbed system $H_0$ into the exact ground state $|\psi_0\rangle$ of 
$H = H_0+H_1$.
Explicitly it is given by \cite{BeckFul88}
\begin{equation}
\Omega = 1 + \lim_{x \to 0} {1 \over {x-(L_0+H_1)}}H_1 \,.
\end{equation}
The brackets $\langle\phi_0|\,...\,|\phi_0\rangle^c$ denote cumulant expectation
values with $|\phi_0\rangle$.
The dot $\cdot$ in Eq. (\ref{KUMKORR}) indicates that the quantity 
inside $(...)^{\cdot}$ 
has to be treated as a single entity in the cumulant formation.
$L_0$ is the Liouville operator corresponding to $H_0$, i.e., 
$L_0 A = [H_0,A]_{-}$.
The relation (\ref{KUMKORR}) can be applied to either weakly or strongly
correlated systems because its use is independent of the operator
statistics , i.e., it is valid for fermions, bosons or spins.

Using Mori-Zwanzig projection technique \cite{Mori} one can derive
a set of equations of motion for the dynamical correlation
functions $G_{\nu\mu}(z)$. Neglecting the self-energy terms 
as explained below it reads:
\begin{equation}
\sum_{\nu} \left( z\delta_{\eta\nu}-\omega_{\eta\nu}
  \right)
  \,G_{\nu\mu}(z)\,\,=\,\,\chi_{\eta\mu} \, .
\label{PROJ_GLSYS}
\end{equation}
$\chi_{\eta\nu}$ and $\omega_{\eta\nu}$
are the static correlation functions and frequency terms,
respectively.
They are given by the following cumulant expressions:
\begin{eqnarray}
\chi_{\eta\nu} \,&=&\,\,
  \langle\phi_0|\,\Omega^\dagger\,B_{\eta}^\dagger B_{\nu}^{\phantom\dagger}\,
                  \Omega\,|\phi_0\rangle^c \, ,
  \nonumber\\
\omega_{\eta\nu} \,&=&\,\,
  \sum_{\lambda}
  \langle\phi_0|\,\Omega^\dagger\,B_{\eta}^\dagger (LB_{\lambda}^{\phantom\dagger})^{\cdot}\,
                  \Omega\,|\phi_0\rangle^c \,\chi_{\lambda\nu}^{-1} \, ,
\label{MATRIX_KUMDEF}
\end{eqnarray}
$\chi_{\nu\mu}^{-1}$ is the inverse matrix of $\chi_{\nu\mu}$. 
These terms describe all dynamic processes within the subspace of the
Liuoville space spanned by the operators $B_{\nu}$.

Next we outline the description of the ground state of the 
$t$-$J$ model at weak doping within the cumulant formalism, 
details have been published recently \cite{VojBeck96}.
$\delta$ denotes the hole concentration away from half filling,
i.e., the system with $N$ lattice sites possesses $M=\delta N$ dopant holes.
The Hamiltonian is decomposed into $H_0$ and $H_1$ as follows:
\begin{eqnarray}
H_{0}\, =\, H_{Ising} \, , \quad 
H_{1}\, =\, H_{t}\, +\, H_{\bot} 
\,.
\end{eqnarray}
The ground state $|\phi_0\rangle$ of the unperturbed Hamiltonian $H_0$ is an
antiferromagnetically ordered N\'{e}el state with $M$ holes. The holes have
fixed momenta ${\bf k}_m$ and are located on the sublattice $\sigma_m$
($\sigma_m = \uparrow,\downarrow$)
\begin{eqnarray}
|\phi_0\rangle\,\, &=&\,\,
  \hat{c}_{{\bf k}_{1}\sigma_{1}}\, \ldots\,\hat{c}_{{\bf k}_{M}\sigma_{M}}
  \, |\phi_{N{\mathaccent 19 e}el}\rangle \,, \nonumber\\
\hat{c}_{{\bf k}\uparrow}\,\, &=&\,\,
 (N/2)^{-1/2} \sum_{i\in\uparrow}\,
  {\rm e}^{i\,{\bf k}{\bf R}_{i}}\,
  \hat{c}_{i\uparrow} \, ,\quad
\hat{c}_{{\bf k}\downarrow}\,\, =\,\,
 (N/2)^{-1/2} \sum_{i\in\downarrow}\,
  {\rm e}^{i\,{\bf k}{\bf R}_{i}}\,
  \hat{c}_{i\downarrow} \,.
\label{GSDEF3}
\end{eqnarray}

Within the cumulant method we employ an exponential ansatz \cite{SchorkFul92} 
for the wave operator $\Omega$, i.e., $\Omega = {\rm e}^S$. 
The operator $S$ describes the effect of the perturbation $H_1$ onto the unperturbed
ground state $|\phi_0\rangle$. 
We are basically interested in the charge dynamics 
of the system.
Therefore we assume that background spin fluctuations can be neglected. 
To treat hole motion processes induced by $H_t$ the concept
of path operators \cite{ShrSig} is used
which leads to the picture of spin-bag quasiparticles \cite{Schrieffer}.
Here we define path concatenation operators $A_{n,\rho}(i)$ where $i$ denotes
a lattice site, $n$ the path length, and $\rho$ the individual path shape:
$A_{n,\rho}(i)$ operating on a hole at site $i$ in the state (\ref{GSDEF3})
moves the hole
$n$ steps away and creates a path or string of $n$ spin defects
attached to the transferred hole.
Note that there is a number $m_n$ of different path shapes for a given 
path length n. For $n=1$ there are $m_1 = 4$ different paths,
for $n=2$ $m_2 = 12$ and so on.
Having defined the excitation operators $A_{n,\rho}(i)$, 
the wave operator $\Omega$ of the cumulant formalism takes the form
\begin{equation}
\Omega\, = \, {\rm exp} \left(\,
  \sum_{n=1}^{n_{max}}\sum_{\rho=1}^{m_n} \lambda_{n,\rho} A_{n,\rho}\, \right) \,,\quad
A_{n,\rho}\,=\,\sum_i A_{n,\rho}(i)
\label{STOEROPDOT2}
\end{equation}
with parameters $\lambda_{n,\rho}$ yet unknown.
Note the additional definiton 
$A_0(i) = {1 \over 2} \sum_\sigma 
\hat c_{i\sigma}^{\phantom\dagger} \hat c_{i\sigma}^\dagger$
which is the only "path" with zero length ($\rho=1$).
The path operators $A_{n,\rho}$ commute with each other because they only
contain spin-flip operators destroying N\'{e}el order.

Following ref. \cite{SchorkFul92} one attains a non-linear set of equations
for the ground-state energy $E_0$ and the coefficients $\lambda_{n,\rho}$:
\begin{eqnarray}
E_{0}\,=\, \langle\phi_0| H \Omega |\phi_0\rangle^c \, , \quad
0    \,=\, \langle\phi_0| A_{n,\rho}^\dagger\, H \, \Omega\, |\phi_0\rangle^c
\label{GLSYSPFADSPINW}
\end{eqnarray}
with $\Omega$ given by (\ref{STOEROPDOT2}).
The cumulant expectation values have to be taken with respect to the unperturbed
ground state (\ref{GSDEF3}). 
The set of equations (\ref{GLSYSPFADSPINW}) together with the additional 
approximation of independent hole quasiparticles leads to a generalized
eigenvalue problem. For details see \cite{VojBeck96}. The used picture of
independent quasiparticles is appropriate for small hole concentrations,
i.e., for a dilute gas of holes moving in an antiferromagnetic background.

For the application of Mori-Zwanzig projection technique one has to choose
a set of relevant operators $B_{\nu}$.
Here we are going to neglect the self-energy terms 
obtained within projection technique, cf. (\ref{PROJ_GLSYS}).
Therefore we need to use a sufficiently
large set of operators $B_{\nu}$ to cover the charge dynamics.
One variable to include is the charge density operator 
$\rho_{\bf k}$ itself. The other variables should provide a coupling between
the ground state $\Omega|\phi_0\rangle$ and all excited states of $H$.
These variables are obtained by applying the perturbation $H_1$
to $\rho_{\bf k}$.
Assuming that the charge dynamics is mainly carried by spin-bag
quasiparticles we can define
\begin{equation}
B_{n,\rho}({\bf k})\,\,=\,\,
  \sum_i e^{{\rm i} {\bf k} {\bf R}_i} \,\,
  A_{n,\rho}(i)\, A_0(i)^\dagger \,\,=\,\,
  \sum_i e^{{\rm i} {\bf k} {\bf R}_i} \,\,
  B_{n,\rho}(i) \, .
\end{equation}
As explained above, $A_{n,\rho} (i)$ is a path operator of length $n$ 
with the path shape $\rho$ acting on a hole at site $i$.
The indices $n,\rho$ replace the general index $\nu$ of the dynamical
variables $B_{\nu}$ in Eqs. (\ref{ARBKORR},\ref{KUMKORR}).
The Fourier-transformed operator $B_{n,\rho}(i)$ couples to a hole 
(at site $i$) and adds the path $A_{n,\rho}$.
So the relevant part of the Liouville space spanned by the operators $B_{n,\rho}$
contains all individual path states connected to the holes.

The first of these operators is the hole density operator itself:
\begin{equation}
B_{0,1}({\bf k}) \,=\, {1 \over 2} \rho_{\bf k}^{hole} \,=\,
  {1 \over 2}
  \sum_{i\sigma} {\rm e}^{{\rm i} {\bf k} {\bf R}_i} \,\,
  {\hat c}_{i\sigma}^{\phantom +} {\hat c}_{i\sigma}^\dagger
\end{equation}
The quantity we are interested in
is therefore the diagonal correlation function $G_{\nu\nu}({\bf k},z)$
with $\nu=(0,1)$.
Note that particle density and hole density are equivalent quantities.

\begin{figure}
\epsfxsize=11cm
\epsfysize=8cm
\hspace{3.5cm}
\epsffile{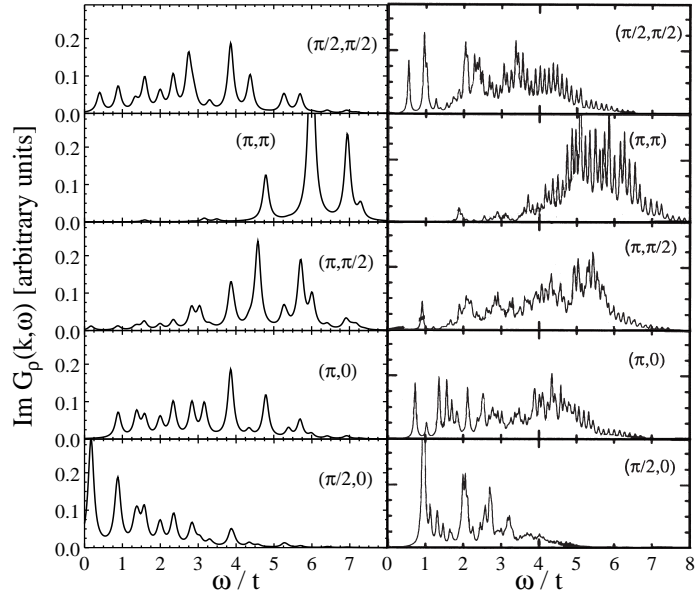}
\vspace{0.5cm}
\caption{Left panel: charge-charge correlation function obtained
from the present calculation for $t/J$ = 2.5, very small $\delta$, 
$n_{max} = 5$, and different momentum transfers, 
plotted with Lorentzians using an artificial linewidth of 0.1 $t$.
Right panel: Exact diagonalization data from \protect\cite{ToHoMae95}
for comparison, $t/J = 2.5$ (see also \protect\cite{KhaHo96}). 
The numerical data have been calculated for a 4 x 4 periodic cluster and a hole 
concentration of $\delta = 25\%$.}
\end{figure}

Our results for the charge-response function ${\rm Im}\,G_\rho({\bf k},\omega)$ 
at $t/J = 2.5$ are shown in Fig.~1. 
The intensity of the spectra is proportional to the hole concentration
$\delta$. 
For small momentum the spectral weight is mainly
concentrated in a peak near $\omega=0$ (with an energy scaling 
with ${\bf v}_F\cdot{\bf k}$ for ${\bf k}\rightarrow 0$).
With increasing momentum we find a transfer of spectral weight 
to higher energies. Calculations for other values of $t/J$ show that the
broad structures observed in the spectra for large momenta scale with $t$,
e.g., the maximum spectral weight in the charge-response function
at $(\pi,\pi)$ remains at energies of about $6t$.
The right panel of Fig.~1. shows exact 
diagonalization data for $\delta = 25\%$ taken from
ref. \cite{ToHoMae95}. We observe a good agreement of the spectra, 
some differences may be either due to the small number of sites in 
the numerical calculations (which leads to an energy gap in
all spectra and other finite-size effects) 
or due to the neglection of relaxation processes in the present 
calculation (such processes would produce finite linewidths).

For the interpretation of the calculated spectra one can consider
the one-hole spectrum of the Hamiltonian within our approximation.
From the diagonalization of the one-hole problem in the subspace
of path operators $A_{n\rho}$ one obtains several bands for the 
spin-bag quasiparticles.
As is well known, the lowest band has a
minimum at $(\pm\pi/2,\pm\pi/2)$ and a bandwidth of about $2J$.
It corresponds to a quasiparticle with s-like symmetry. 
The spin-bag states in the higher bands have nodes in the coefficients 
$\lambda_{n,\rho}$. 
The low-energy peak in the spectra (visible at momentum transfer
$(\pi/2,0)$ ) corresponds to excitations within
the lowest quasiparticle band whereas the high-energy part arises from
transitions to higher bands.

\begin{figure}
\epsfxsize=12cm
\epsfysize=5.5cm
\hspace{1.5cm}
\epsffile{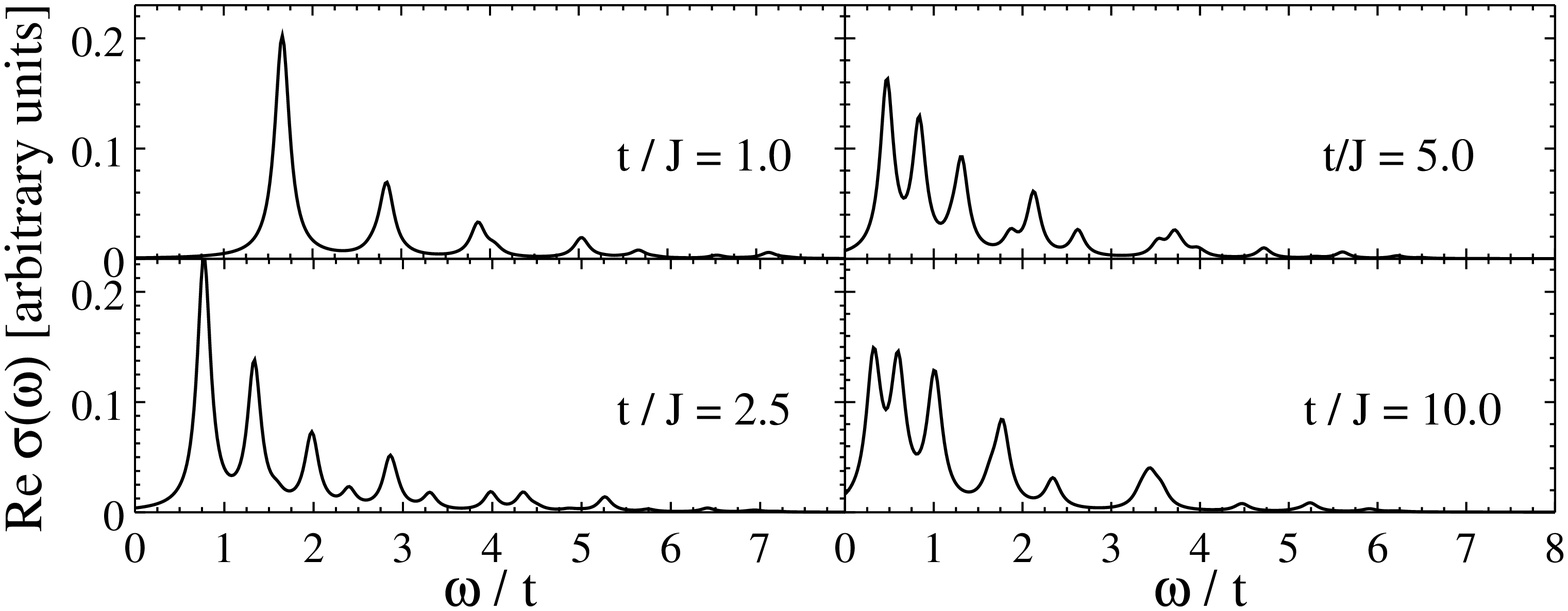}
\caption{Optical conductivity for small hole concentration and different 
$t/J$. The spectra have been calculated from the charge-charge correlation 
function using (\protect\ref{OPTCON}).}
\end{figure}

Using Eq. (\ref{OPTCON}) one can deduce the optical conductivity from
the calculated charge response function.
Results for ${\rm Re}\,\sigma(\omega)$ are displayed in Fig.~2. 
They show again good agreement with the numerical results
from ref. \cite{EdWroOh96} for $\delta = 12.5\%$.
The main peak (at low energy) scales with $J$, it is located at
$\approx 1.7 J$. It arises from the 
transition between the first and second quasiparticle band, i.e., from
the s-like groundstate to a p-like state. So our calculation
supports the discussion given in ref. \cite{EdWroOh96} where only these
two states have been considered in a simplified analytical calculation
for the main peak.
We want to emphasize that taking into account all string states
(up to a truncation length) as done in the present work reproduces not
only this main peak but also the incoherent continuum found at energies
up to $6t$. 
The low-energy peak corresponding to $\omega \approx$ 0.3 eV for
$J =$ 0.15-0.2 eV which was
also found in numerical studies of the $t$-$J$ and Hubbard models
is supposed to coincide with the mid-infrared peak observed in
optical spectra of high-$T_c$ superconductors (see e.g. \cite{Dagotto94}).
Within the present calculation we do not obtain a Drude peak 
$D\,\delta(\omega)$ since Eq.
(\ref{OPTCON}) is valid for non-zero frequency only, i.e., it does
not include the diamagnetic part of the current.
The different scaling behaviour of $G_\rho(\bf k,\omega)$ (for
large $\bf k$) and $\sigma(\omega)$ with $t/J$ arises from the internal
structure of the spin bag forming the quasiparticle.
Details will be published elsewhere \cite{VojBeck97}.

The agreement of our data with the exact diagonalization results
\cite{ToHoMae95,EdWroOh96} is
remarkable because these numerical calculations are done for rather 
large hole concentrations (e.g., 25\%) where only short-range 
magnetic order is present in the high-$T_c$ materials.
In contrast, the present calculations are based on magnetic long-range order. 
Thus we conclude that whether the quasiparticles move in a long-range
ordered background or not does not have an essential influence on the
charge response of the system.
The dynamics is mainly determined by the local antiferromagnetic order in the 
vicinity of the hole quasiparticles \cite{AFDop}, i.e., the magnetic correlation length 
has to be of the order of the quasiparticle size. 

One should note that a recent slave-boson approach \cite{KhaHo96}  
to the charge dynamics in the $t$-$J$ model also reproduces 
some key features of the charge response function, namely a
peak at small energy for small ${\bf k}$ and a broad featureless continuum
at energies of several $t$.
The theory presented there is based on a Fermi liquid picture and
completely neglects antiferromagnetic correlations in the ground state.
However, in the high-$T_c$ materials short-range correlations are present 
also at higher hole concentrations. 
We believe them to be important for the structure of the spectra.

Summarizing, we
conclude that the charge dynamics in the weakly doped $t$-$J$
model can be well explained within the picture of 
independent spin-bag quasiparticles and their excitations.
The present calculation may serve as a basis for the investigation of
the charge dynamics in more realistic models for the cuprate superconductors,
as e.g. the three-band Hubbard model.

\end{document}